\documentclass[conference]{IEEEtran}
\ifCLASSINFOpdf
\else
\fi
\makeatletter
\def\endthebibliography{%
  \def\@noitemerr{\@latex@warning{Empty `thebibliography' environment}}%
  \endlist
}
\makeatother

\usepackage{hyperref}
\usepackage{graphicx}
\usepackage{longtable}
\usepackage{multirow}
\usepackage{blindtext,graphicx}
\usepackage[absolute]{textpos}
\usepackage{color}

\newcommand{\todo}[1]{}
\newcommand{\lili}[1]{}
\newcommand{\Xin}[1]{}
\renewcommand{\todo}[1]{{\color{red} TODO: {#1}}}
\renewcommand{\lili}[1]{{\color{blue} LI: {#1}}}
\renewcommand{\Xin}[1]{{\color{red} Xin: {#1}}}
\begin{document}
\begin{textblock}{18}(1,1)
Author pre-published version. The paper is accepted for publication in IEEE/ACM 43nd International Conference on Software Engineering: Software Engineering in Society (ICSE-SEIS), 2021.
\end{textblock}
%

\title{A First Look at Human Values-Violation \\ in  App Reviews}


%
\author{\IEEEauthorblockN{Humphrey O. Obie\IEEEauthorrefmark{1},
Waqar Hussain\IEEEauthorrefmark{1},
Xin Xia\IEEEauthorrefmark{1}, 
John Grundy\IEEEauthorrefmark{1},
Li Li\IEEEauthorrefmark{1},
Burak Turhan\IEEEauthorrefmark{1},
Jon Whittle\IEEEauthorrefmark{2},
Mojtaba Shahin\IEEEauthorrefmark{1}}
\IEEEauthorblockA{\IEEEauthorrefmark{1}Monash University, Melbourne, Australia}
\IEEEauthorblockA{\IEEEauthorrefmark{2}CSIRO's Data61, Melbourne, Australia \\
\{humphrey.obie, waqar.hussain, xin.xia, john.grundy, li.li, burak.turhan, mojtaba.shahin\}@monash.edu, jon.whittle@data61.csiro.au}}

%


\maketitle

\begin{abstract}
Ubiquitous technologies such as mobile software applications (mobile apps) have a tremendous influence on the evolution of the social, cultural, economic, and political facets of life in society. Mobile apps fulfil many practical purposes for users including entertainment, transportation, financial management, etc. Given the ubiquity of mobile apps in the lives of individuals and the consequent effect of these technologies on society, it is essential to consider the relationship between human values and the development and deployment of mobile apps. The many negative consequences of violating human values such as privacy, fairness or social justice by technology have been documented in recent times. If we can detect these violations in a timely manner, developers can look to better address them. To understand the violation of human values in a range of common mobile apps, we analysed 22,119 app reviews from Google Play Store using natural language processing techniques. We base our values violation detection approach on a widely accepted model of human values; the Schwartz theory of basic human values. The results of our analysis show that 26.5\% of the reviews contained text indicating user perceived violations of human values. We found that benevolence and self-direction were the most violated value categories, and conformity and tradition were the least violated categories. Our results also highlight the need for a proactive approach to the alignment of values amongst stakeholders and the use of app reviews as a valuable additional source for mining values requirements.

\end{abstract}


%
\IEEEpeerreviewmaketitle

\section{Introduction}

Ubiquitous technologies have a tremendous influence on the evolution of the social, cultural, economic, and political facets of life in contemporary society, and their effects cannot be ignored \cite{Mohamed:2020,Agre:1997}. Smartphones with the accompanying mobile software applications (mobile apps) are an example of such ubiquitous technologies. Mobile apps fulfil many practical purposes for users, ranging from entertainment (e.g., YouTube), to transportation (e.g., Google Maps), and financial management (e.g., banking apps). Currently, there are more than 2 billion smartphone users in the world, and 5 million apps available to download from both the Apple App Store and the Google Play Store \cite{Appstat:2020}. Given the ubiquity of mobile apps in the everyday life of users and the consequent effect on society, it is pertinent to consider the relationship between human values and the development and deployment of mobile apps, and how mobile apps, in turn, embody human values.

There are various stakeholders involved in the development and deployment of software systems (including mobile apps), e.g., business analysts, developers, and end-users. All of these stakeholders have different  \emph{human values} considerations that may or may not be explicitly acknowledged. 
Others have suggested that systems have no conscience and therefore embody the values of their creators, i.e., software engineers \cite{Lennox:2020}. In a recent work on measuring values in software engineering, Winter et al. presented three software engineer prototypes, i.e., abstract types of software engineers, and their preferred ranking of values \cite{Winter:2018}. The second software engineer prototype in their study ranked “being an honest and trustworthy colleague” and the “achievement of high quality” software lower than other factors such as the “software being commercially successful” and the “software influencing the end-user”. If the values of all stakeholders -- especially those of the end-users -- are not properly articulated, documented, and agreed upon, it is probable that the values of the software engineers (e.g., values held by the prototypes in \cite{Winter:2018}) would inadvertently be represented in the resulting system.

There have also been calls to better align technology with the values of users in other areas such as artificial intelligence (AI). For example, the Asilomar principle on human values states that, ``AI systems should be designed and operated so as to be compatible with ideals of human dignity, rights, freedoms, and cultural diversity'' \cite{Asilomar:2020}. There have also been some recent and isolated efforts on values-based software and requirements engineering \cite{Perera:2020,Thew:2017,Ferrario:2016,Aldewereld:2015}.

Although software engineering practice and research have captured well-known values such as privacy, security, and accessibility, little attention has been paid to broader human values such as conformity and self-direction in software engineering, especially in mobile apps \cite{Shams:2020,Perera:2020}. Thus there is the need for studies to provide an understanding of the interplay between human values and mobile technologies, and the negative consequences of violating human values in mobile apps. 

Many examples of the negative consequences of violating different human values in technology have been reported in the media. Recently, Robodebt, an automated debt recovery tool, distressed thousands of Australians on social welfare, some to the point of suicide, by issuing them inaccurate debt notices \cite{Nine:2020,ACS1:2020}. This led to a public outcry and the government having to waive or refund over half a million dollars to the affected people \cite{ACS2:2020}. Besides, in the Facebook-Cambridge Analytica scandal, Facebook was held responsible for allowing Cambridge Analytica to collect over 50 million users’ data without their consent \cite{Guardian:2018}, resulting in a 5 billion dollars fine by the Federal Trade Commission (FTC) \cite{FTC:2020}, and a loss of 119 billion dollars in its market stock value \cite{Rupert:2020}. As another example, Instagram was partly blamed for the death of a British teenager \cite{Angus:2020}, leading them to promptly remove images related to suicide and self-harm \cite{Sarah:2020}. These examples highlight just some deleterious effects of technology on contemporary society and the need to take into account ethical and human values consideration in software engineering practice because “values are the key to unlocking ...the enormous dangers of contemporary technologies” \cite{Barn:2015,Goguen:2004}.

To understand the violation of human values in mobile apps, we leverage information contained in users’ apps reviews. App reviews are a valuable source of information, ideas, and requests from users \cite{Guzman:2014,Disorbo:2016}, and we conjecture that app reviews are capable of reflecting human values in mobile apps. Manually analysing feedback from users is a time-consuming process since popular apps typically receive thousands of reviews daily. As a result, automated techniques have been proposed in the literature to reduce the efforts needed for analysing reviews.
In this work reported here, we analysed a total of 22,119 app reviews from 12 apps from Google Play Store to identify violations of human values in mobile apps using  natural language processing (NLP) techniques. Our approach combines sentiment analysis, app feature extraction, and a human values dictionary of keywords. Our approach achieves a recall of 0.83, precision of 0.69, and F-measure of 0.75. We base our values detection approach on a widely accepted model of human values, the Schwartz theory of basic human values \cite{Schwartz:1992,Schwartz:2012}. The result of our analysis shows that 26.5\% of the reviews contained violations of human values. Benevolence and self-direction were the most violated value categories, while conformity and tradition were the least violated value categories. The results also relate certain app features to specific values violation. Furthermore, this study is presented as one of the early steps in understanding the nexus between human values and mobile apps.

This work makes the following key contributions:
\begin{itemize}
    \item We present evidence for the reflection of human values and their violations in app reviews as a proxy for mobile apps.
    \item The results of our analysis of app reviews highlights the most prevalent categories of violation of human values.
    \item We present a set of recommendations to enable better support of the key end user human values in mobile apps.
\end{itemize}


\section{Motivation}
\label{sec:motivation}
Consider a mobile app developed to provide paid parking services to institutions such as city councils, universities, hospitals, parks, etc., and users sign up to the app for a more convenient method of parking instead of the use of the traditional pay machine. The app supports tasks such as a pro-rated payment system that allows users to pay for only the time they park; allows users to see how much time they have left in a parking session and get reminders to know when a session is about to expire; and also allows them to park now and pay later. 

Unfortunately, while well-intentioned, the app violates many of the human values of its different users. Many users are not comfortable with revealing their locations and reluctant to give a parking app access to their photos -  an issue related to the value of \emph{privacy}. For example, a user writes, \emph{“There is no way I’m giving your app my location… or access to my photos”}. Other users find that the app does not fulfil its primary purpose of \emph{helping} them find and pay for a parking space. Due to certain physical or cognitive disability, the app \emph{induces anxiety} in them and decreases their feeling of \textit{independence}. An example user review describes this situation: \emph{“I wanted to find a hospital parking place... but this app finds none so I'm afraid it is completely useless for me. I’m disabled and cannot walk a long way to pay for a parking ticket so I thought the app was going to be very helpful.}” Moreover, some users find the lack of notification about auto-sign up for premium subscription and their inability to cancel their subscription and remove their credit card details as \emph{dishonest} behaviours by the app providers. Several reviews mentioning this relate this to the values of \textit{honesty} and \textit{self-direction}. A representative review captures this: \emph{“...It is unscrupulous about signing you up to a subscription when you are skipping past the in-app ads. It is not made clear once you have subscribed and there is no way of cancelling it through the app.”}

Such values-violating app defects can remain unfixed for a long time if there is no driver to prioritise them. Misunderstood user requirements, cost-minimising and profit-maximising business and design decisions, and software bugs can all easily result in violations of human values. These defects result in poor take-up of apps, confused or unhappy users, organisational reputational damage and lost customers. Hence our work aims to contribute an understanding of the violations of human values in mobile apps by detecting these violations in app reviews. We aim to help developers to better understand and identify these end user human values violations in their apps with a view to being able to both fix them and avoid them in future. Following this aim, we guided our study with the following questions:

\begin{enumerate}
    \item [RQ1] What are the most common human values violations perceived by app users as documented in their mobile app reviews?
    \item [RQ2] Which reviews indicating human values violations by a mobile app are the most supported by other end users?
    \item [RQ3] Are app description features related to specific human values violations?
\end{enumerate}

\section{Human Values}
\label{sec:humanvalues}


Human values -- defined as the  ``guiding principles of what people consider important in life'' \cite{Cheng:2010} -- have been well studied in the social sciences, resulting in various concepts and theories of human values \cite{Whittle:2019}.

Rokeach postulates that human values determine human behaviour and attitude and proposed the Rokeach value scale \cite{Rokeach:1973}. Rokeach characterised 36 human values into 2 main categories - 18 terminal values which refer to goals in life; and 18 instrumental values which refer to modes of conduct.


In their work on the perception of values by future professionals, Parashar et al. surveyed students on their impressions of values that currently exist and those that ought to exist in society \cite{Parashar:2004}. They contend that there are two fundamental levels of human values; micro-level and macro-level. Values at the individual micro-level are internalised standards that reconcile a person’s needs with the demands of social life, whereas values at the macro-level of cultural practices represent shared understanding that gives meaning, order, and integration to social living.

The fundamental theory of values posits values as a guide for actions and a vehicle for expressing need \cite{Gouveia:2014}. However, the most widely accepted theory of human values is the Schwartz theory of basic human values \cite{Schwartz:1992,Schwartz:2012}.

The Schwartz theory of basic human values categorises 58 human values into 10 categories \cite{Schwartz:1992,Schwartz:2012}. This theory is based on a survey carried out in several countries and covers different factors including age, gender, cultural practices, and geography. The 10 categories and their associated value items represent distinct motivational value orientations general to all cultures and relate to the fundamental need of human existence (See \autoref{tab:values}). 

Furthermore, the Schwartz theory of basic human values have gained wide acceptance in other disciplines beyond the social sciences and has seen adoption in computer science and software engineering research \cite{Shams:2020}. Studies in software engineering focusing on operationalising human values have applied Schwartz’s theory. For example, the values Q-sort measures human values at different levels of abstraction in software engineering \cite{Winter:2018}, while the values-hub aims to integrate human values into software design patterns \cite{Hussain:2018}. Thus, in this work, we utilise the Schwartz theory based on its wide adoption and application.
 
\begin{table}
\scriptsize
\centering
\caption{Value categories and descriptions \cite{Schwartz:2012}.}
\label{tab:values}
\begin{tabular}{lp{6cm}} 
\hline
Value Category & Description (motivational goals)                                                                                          \\ 
\hline
Self-direction & Independent thought and action - choosing, creating, exploring                                                            \\ 
\hline
Stimulation    & Excitement, novelty, and challenge in life                                                                                \\ 
\hline
Hedonism       & Pleasure or sensuous gratification for oneself                                                                            \\ 
\hline
Achievement    & Personal success through demonstrating competence according to social standards                                           \\ 
\hline
Power          & Social status and prestige, control or dominance over people and resources                                                \\ 
\hline
Security~      & Safety, harmony, and stability of society, of relationships, and of self                                                  \\ 
\hline
Conformity     & Restraint of actions, inclinations, and impulses likely to upset or harm others and violate social expectations or norms  \\ 
\hline
Tradition      & Respect, commitment, and acceptance of the customs and ideas that one's culture or religion provides                      \\ 
\hline
Benevolence    & Preserving and enhancing the welfare of those with whom one is in frequent personal contact                               \\ 
\hline
Universalism   & Understanding, appreciation, tolerance, and protection for the welfare of all people and for nature                       \\
\hline
\end{tabular}
\end{table}


\section{Methodology}
\label{sec:approach}


The aim of our approach is to automatically detect human values violations manifesting in user app reviews, associate these violations to specific app features, and then provide a better understanding of the violations of human values for mobile app software engineers. 

To achieve our aim and answer our research questions, we applied natural language processing (NLP) techniques to a corpus of mobile app reviews for a set of popular mobile apps.
Firstly, we extract app descriptions and user app reviews from selected apps including the review text, star rating, and the number of likes for each review. Then we carry out preprocessing steps on the review text to prepare it for our NLP values violation detector and app description features extraction. Next, we extract app description features from the app descriptions, and then we apply lexical sentiment analysis to the review text to detect user sentiments. Finally, we apply our NLP detector in combination with the sentiment results and extracted app features to detect values violation in the review text and their associated app features. 
We describe the details of each step in the following subsections.




\subsection{Data Collection}
To discover human values violations prevalent in app reviews, we analysed reviews from 12 popular apps using our values-violation detection approach. All 12 apps are available in the Google Play store. The selected apps were chosen to cover different audiences and age-groups, with different expectations and interactions with technology, and to see how human values violations by the apps may be represented in different app user reviews. \autoref{tab:dataset} shows the details of the apps. The total number of reviews collected was 22,607. We collected the most recent 2,000 reviews from each app; however, only 607 reviews were available for the Cellopark app at the time of data collection. After discarding non-informative reviews, i.e., reviews with less than three tokens, we were left with a total of 22,119 reviews used for our analysis.

\begin{table}
\scriptsize
\centering
\caption{Overview of the dataset.}
\label{tab:dataset}
\begin{tabular}{lll} 
\hline
App Name      & Category         & \#Reviews  \\ 
\hline
Pinterest     & Social           & 2000       \\
TrainingPeaks & Health  Fitness  & 2000       \\
Minecraft     & Games            & 2000       \\
Monopoly      & Games            & 2000       \\
PicsArt       & Photography      & 2000       \\
Any.do        & Productivity     & 2000       \\
Telegram     & Communication    & 2000       \\
Tripadvisor   & Travel           & 2000       \\
PayByPhone    & Auto  Vehicles   & 2000       \\
Cellopark     & Maps  Navigation & 607        \\
Tiktok        & Social           & 2000       \\
CBA            & Finance          & 2000       \\
\hline
\end{tabular}
\end{table}

\subsection{Preprocessing}
\label{subsec:prep}
In this phase, we extract and preprocess app reviews from the Google Play store (app distribution platform). These user reviews include the following attributes: review text, star rating, and the number of likes for each review.
Then we preprocess the extracted app reviews in the following steps:
k
\subsubsection{Misspelt Words}
It is common for users to make typographical errors and misspell words while leaving app reviews since most users write their reviews on mobile devices. This is mainly due to the limited sizes of mobile devices, and the lack of a dedicated physical keyboard \cite{Phong:2015}). For example, in the reviews, the word “pretty” is misspelt “pritty”, and “share” as “sharr”.
To identify and fix these spelling errors, we utilise the \textit{autocorrect} spell checker library\footnote{https://github.com/fsondej/autocorrect}. The autocorrect library implements the Levenshtein distance \cite{Doan:2012} algorithm to search for word permutations within an edit distance of 2 and then compares the results to known words in a word frequency list.

\subsubsection{Stopwords Removal and Stemming}
Making use of the NLTK library\footnote{https://www.nltk.org/}, we remove common English stopwords, e.g., this, is, a, etc. These stopwords are frequent in user reviews but do not provide useful information.
Moreover, information retrieval systems typically utilise stemming to improve their searching capabilities \cite{Phong:2015}. We use the English Snowball stemmer to reduce the inflection in words to their stems.

\subsection{App Description Features Extraction}
To extract app features from app descriptions, we adopted the SAFE approach proposed by Johann et al. \cite{Johann:2017}. SAFE, the current state-of-the-art in app feature extraction, is a rule-based method for extracting features from both app descriptions and user app reviews. The advantage of SAFE is that it does not require large training data and instead relies on manually curated part-of-speech patterns and sentence patterns that are frequently used in text to refer to app features. The output from the app features extraction is a list of fine-grained features, e.g., file viewer, collage maker, etc. Two of the authors manually verified the feature extraction output to make sure they are valid app features.

\subsection{Values-Violation Detection}
The key aim of our approach is to automatically detect values violations in apps from issues reported in user app reviews. In this phase, we identify likely end user human values violations using NLP techniques.

\subsubsection{Sentiment Analysis}
Sentiment analysis is the process of detecting an affect or mood in a text by assigning a quantitative value in a text corpus \cite{Panichella:2015}. Three possible classes of sentiment intensity can be assigned: positive, negative, and neutral. For analysing the sentiments expressed in the reviews, we employ the VADER sentiment analysis model \cite{Hutto:2015}, a less resource-consuming lexicon and rule-based model that is attuned to sentiments expressed in text corpus such as user reviews. The VADER model computes a normalised weighted composite score (compound score) by summing the valence scores of each word in the lexicon, adjusted to the grammatical and syntactic rules, and then normalised to be between $-1$ (most negative) and $+1$ (most positive). 

The generally accepted thresholds for the compound score, $x$, for each class of sentiment intensity in the literature are as follows: positive ($x >= 0.05$), neutral ($x >-0.05  \land x < 0.05$), and negative ($x <= -0.05$).

\subsubsection{Definition of Values Dictionary}
For our definition of human values, we refer to Schwartz’s theory of basic human values \cite{Schwartz:1992,Schwartz:2012}. There are 58 human values categorised into 10 categories in Schwartz’s theory. We argue that exploring appropriate keywords associated with Schwartz’s values terminology provides important information to better support understanding human values reflected in users app reviews.
Two authors of the paper created and validated a dictionary of human values consisting of 50 human values using Schwartz’s values terminology. 

This dictionary consists of these values terminology and their associated synonyms and antonyms. To make the dictionary thorough and comprehensive, we used entries from an online thesaurus\footnote{https://www.thesaurus.com/} and Merriam-Webster dictionary\footnote{https://www.merriam-webster.com/}. To address coherency, on the other hand, we only included entries that are semantically and contextually related to the values terms \cite{Wu:1994}. This way, every value term has semantically relevant synonyms and antonyms. We initially used the WordNet lexical database \cite{Miller:1995} to provide synonyms and antonyms for the values dictionary. However, WordNet is limited in terms of the number of entries and had poor contextual relevance to some of the values terms, e.g., the value term, “helpful” has no synonym in WordNet, while it lists “chair” as a synonym for the value term, “moderate”.

We hypothesise that on top of a well-curated dictionary of human values, it is feasible to create an NLP classifier capable of detecting human values violation in user reviews. We stem the entries of the values dictionary using the Snowball stemmer, similar to the preprocessing step described in \autoref{subsec:prep} above.

\subsubsection{Truthset Definition}
Having created a comprehensive and contextually relevant values dictionary, two researchers with extensive experience in value-based software engineering manually labelled a truthset ($T_Test$) consisting of 709 reviews dataset. They did a pilot round, labelling 143 reviews each, followed by randomly picking 20 reviews labelled by the other reviewer to check labelling/coding consistency. The main objective was to calibrate decisions and use this to refine the common protocol of identifying violation of values and mapping it to Schwartz value categories. The coders resolved their differences through discussions and calibrating the labelling process. The actual study was then performed on 709 reviews dataset divided between the researchers. After dataset  labelling, a representative sample from each researchers labelled data was reviewed by the other to identify and resolve differences as done in the pilot round. This $T_Test$ dataset forms our oracle for evaluating our NLP approach.


\subsubsection{Automatic Values-Violation Detection}
Using the values dictionary and sentiment analysis output in the previous substeps, we built an NLP detector to automatically detect values violation in user reviews and assign a review text to one or more Schwartz values items. We calculate the probability of a review to contain one or more values-violation as contained in the values-dictionary. Formally, let $R$ be a review and $V$ be a value keyword; let $T_V$ be the number of tokens in $R$ that appear in the $V$-related values-dictionary defined above; let $T_R$ be the total number of token in $R$, the NLP detector calculates the probability that $R$ contains value $V$ as $P_{(R,V)} = \frac{T_V}{T_R}$.

To prevent instances where reviews are inaccurately assigned to a value $V$ because just one $V$-related keyword is contained in $R$, the NLP detector assigns to $R$ all the value keywords for which $P_{(R,V)}$ is greater than $0.05$ (i.e., keywords make up at least 5\% of a review). A value violation is assigned if $P_{(R,V)} >= 0.05$ $and$ the sentiment analysis output is negative or neutral (i.e., sentiment compound score, $x < 0.05$).


We evaluated the effectiveness of our approach by comparing the values-violations labels assigned by the NLP detector to the truthset defined in the previous step. We utilised the generally accepted metrics for evaluation in the information retrieval field, i.e., precision, recall, and F-measure. Our approach detected values-violation in the review text with a recall of 0.83, precision of 0.69, and F-measure of 0.75.








\subsection{Threats to Validity}
\emph{Threats to Construct Validity:} 
This is related to the creation of our truthset. Although the labelling of the data was done by two researchers with experience in values-oriented software engineering, it is still a subjective process.
The researchers who labelled the dataset handled all disagreements by discussion between themselves and calibrating their decisions to refine a common protocol, and in situations where there are differences in the assigned category, they brought it to the attention of the other researcher and came to a consensus on it. 




\emph{Threats to Internal Validity:} 
This threat relates to detection errors that could affect our results. Our analysis made use of our NLP values violation detector after achieving an F1-measure of 0.75 on the truthset. Since the NLP detector is not 100\% error-free, there may be false positives (no value violations in reality but the data is labelled otherwise) and false negatives (violations in reality but the detector miss them). 

\emph{Threats to External Validity:}
This threat relates to the apps chosen for our analysis and how it may affect the generalisability of our results. Out of the many apps available on the Google Play store, we chose 12 apps and analysed their reviews, and these might not be representative. To account for this issue, we chose the 12 apps from 11 different categories covering different audiences and age groups, with different expectations and interaction with technology.

\section{Results}
\label{sec:results}

\subsection{RQ1: What are the most common human values violations perceived by app users as documented in their mobile app reviews?}
\label{sec:reflect}

Reviews are a good source for mining information. Since these app reviews are written by users with various human values considerations (whether implicit or explicit), we conjecture that the reviews will contain information representative of (some of) their values. As described above, we postulate that users will report values violations, implicitly or explicitly, in their reviews in various ways.

\begin{figure}[!ht]
		\centering 
		\includegraphics[width=0.75\linewidth]{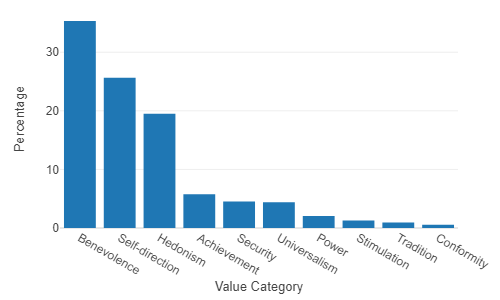}
		\caption{Percentage of violated values category.
		}
		\label{fig:valuecategory}
	\end{figure}

We found that many of the app reviews contain indicators of perceived human values violations. The average number of violations across all 12 apps is 487.6. See \autoref{tab:categoryresult} for a breakdown of the results of values-violations based on Schwartz's values categories for the different apps. Out of the 22,119 app reviews analysed using our values-violation detection technique, there were a total of 5,851 (26.5\% of total reviews) reviews indicating an end user perceived human values violation by the reviewed app. As shown in \autoref{fig:valuecategory}, out of the 5,851 values violations, benevolence (35.3\%), and self-direction (25.6\%), are the most violated values categories, while tradition (0.92\%) and conformity (0.56\%) are the least violated categories.

\autoref{tab:hugetable} shows the human values violations at a finer detail, i.e., value items. Helpfulness, pleasure, and curiosity rank amongst the top 3 human values violations detected in the app reviews, while Obedience and Influential are among the least violated human values. 






\begin{table*}
\scriptsize
\centering
\caption{Value categories, value items and example reviews.
}
\label{tab:hugetable}
\begin{tabular}{|p{1.2cm}|p{1.5cm}|p{0.4cm}|p{13cm}|} 
\hline
\textbf{Value Category }                                             & \textbf{Value Item  }               & \textbf{$f$ }& \textbf{Example of Reviews }                                                                                                                                                                                                                                                                                                                      \\ 
\hline
\multirow{7}{*}{Self-direction}                             & Freedom                    & 56        & \begin{tabular}[c]{@{}p{13cm}@{}}The most glaring issue is that you are \textbf{\textit{confined}} to the app is predetermined reminder intervals...No option to create your own time intervals... They say this app is customizable, as long as it is the developers choice of customizations. look elsewhere if you want total customization.\\ \end{tabular}  \\ 
\cline{2-4}
                                                            & Creativity                 & 83        & It \textbf{\textit{simply}} tries to fix a problem that never existed in the first place, while at the same time adding so many more in the form of petty technical issues and glitches.                                                                                                                                                                   \\ 
\cline{2-4}
                                                            & Independence               & 62        & I do not want a planner app that tries to \textbf{\textit{control}} my phone and tell me what to do next, etc.                                                                                                                                                                                                                                             \\ 
\cline{2-4}
                                                            & Privacy                    & 103       & I just downloaded but after reading \textbf{\textit{privacy}} policy I prefer going thru Google official one and app like calendar required lot of access, and personally do not wanna bleach my data like this                                                                                                                                            \\ 
\cline{2-4}
                                                            & Choosing own goals         & 397       & Is there no way to \textbf{\textit{assign}} a specific color to an event? I like to group my meetings by color. If I cannot do that, the app is not useful for me                                                                                                                                                                                          \\ 
\cline{2-4}
                                                            & Curiousity                 & 792       & No help to be had. No tutorials available. After 3 days I could not even \textbf{\textit{find}} out how to delete notes.                                                                                                                                                                                                                  \\ 
\cline{2-4}
                                                            & Self-respect               & 7         & Inability to transfer between accounts at will have lead to embarrassing situations where friends and family have had to pay. \textbf{\textit{Humiliating}}!                                                                                                                                                                                            \\ 
\hline
\multirow{3}{*}{Stimulation}                                & Excitement in life         & 36        & Functional but unbelievably \textbf{\textit{dull}}. Stops you being unproductive by \textbf{\textit{boring}} you to death within 5 minutes. Get a pen and a piece if paper and make a list; it will be far more thrilling.                                                                                                                                                   \\ 
\cline{2-4}
                                                            & A varied life              & 14        &  It is difficult to tailor your experience to your likings, and the app finds all of one thing forever and ever rather than an \textbf{\textit{assortment}} of ideas.                                                                                                                                                                                     \\ 
\cline{2-4}
                                                            & Daring                     & 26        & what happens, if phone crashes in the middle... and does not automatically save. So I'm not really sure, if I want to use it, because I'm \textbf{\textit{afraid}}.                                                                                                                                                                                                                   \\ 
\hline
\multirow{3}{*}{Hedonism} & Pleasure                   & 876       & This is so \textbf{\textit{frustrating}}!!! I set a event for someone is birthday then a event a month ago before the birthday to not forget to buy the stuff they want and nothing shows up!                                                                                                                                             \\ 
\cline{2-4}
                                                            & Self-indulgent             & 7         & Please make notification control as in the Telegram X. So I can disable pop-up in private chats category, and never see it again. I'm got so tired of doing this for every new chat, that I switched to more \textbf{\textit{restrained}} X version.                                                                                                      \\ 
\cline{2-4}
                                                            & Enjoying life              & 259       & \begin{tabular}[c]{@{}p{13cm}@{}}I'm just not digging the functionality. Trying to set a custom date and time is a complete \textbf{\textit{pain}}\\ \end{tabular}                                                                                                                                                    \\ 
\hline
\multirow{5}{*}{Achievement}               & Ambitious                  & 15        & If I knew there was no grind, I would not have bought it. No events, nothing to \textbf{\textit{strive}} for...So much potential ... \#fail                                                                                                                                                                                                                 \\ 
\cline{2-4}
                                                            & Influential                & 3         & I always see so many fake accounts of \textbf{\textit{famous}} tiktok Stars fake account makers is videos from somewhere else and post to their fake accounts and get million of likes and followers by misleading the people.                                                                                                                     \\ 
\cline{2-4}
                                                            & Capable                    & 29        & Low \textbf{\textit{clumsy}} registration. Could not pay for parking using the app.                                                                                                                                                                                                                                                      \\ 
\cline{2-4}
                                                            & Successful                 & 132       & I understand bugs happen but I'm \textbf{\textit{losing}} the value I paid... I will give 5 stars once my money does not feel wasted.                                                                                                                                                                                                \\ 
\cline{2-4}
                                                            & Intelligent                & 158       & Perhaps I am just \textbf{\textit{stupid}}. Perhaps it does not work with my phone. Uninstalled and we will try something else to work with my Google assistant.                                                                                                                                                                                           \\ 
\hline
\multirow{4}{*}{Power}                                      & Wealth                     & 79        & \begin{tabular}[c]{@{}l@{}}Its very difficult to close your account when you are oversea. And they continue adding to your \textbf{\textit{debt}} even if you are not using it.\\ \end{tabular}                                                                                                                                                            \\ 
\cline{2-4}
                                                            & Authority                  & 18        & I like it and all that but the things that annoy me is people who have had if for a little while get more \textbf{\textit{clout}} than us who have had it from its musically days                                                                                                                                                                          \\ 
\cline{2-4}
                                                            & Preserving my public image & 10        & When someone is reporting a user, tiktok should investigate instead of hiding the videos of the bad user. There are many trollers in tiktok and they are \textbf{\textit{tarnishing}} the girls image and \textbf{\textit{reputation}} as well.                                                                                                                              \\ 
\cline{2-4}
                                                            & Recognition                & 13        & I feel as the monopoly \textbf{\textit{franchise}} I did not expect they would allow their name to be tarnished this way. Disappointing to say the least.                                                                                                                                                                                                  \\ 
\hline
\multirow{6}{*}{Security}                                   & National security          & 56        & What do you want us to do, telegram? Use \textit{\textbf{unofficial}} clients and compromise our \textbf{\textit{security}}? 1 star too is too much to be given                                                                                                                                                                                                              \\ 
\cline{2-4}
                                                            & Family security            & 67        & \begin{tabular}[c]{@{}p{13cm}@{}} My \textbf{\textit{partner}} and I tried sharing a grocery list and some \textbf{\textit{household}} too is, but it would refuse to add things sometimes, or not be in sync, so we would end up with duplicate groceries. Now I have a year is worth of salt and an upset \textbf{\textit{boyfriend}} who had to go back to the store for.\\ \end{tabular}          \\ 
\cline{2-4}
                                                            & Sense of belonging         & 8         & I like many others have reported plenty of videos in regards to \textbf{\textit{racism}} (including blackface), violence toward LGB+ community, rape, pedophilia etc and like clockwork I always get the same response - "We found that the reported content does not violate our Community Guidelines"                                                    \\ 
\cline{2-4}
                                                            & Social order               & 12        & Fool game, but no records or \textbf{\textit{rankings}}. People start off with bad rolls and just quit...                                                                                                                                                                                                                                                  \\ 
\cline{2-4}
                                                            & Healthy                    & 83        & Cause \textbf{\textit{anxiety}} when need to update the app last minute.                                                                                                                                                                                                                                                                                   \\ 
\cline{2-4}
                                                            & Clean                      & 29        & too many emails. I do not want more \textbf{\textit{clutter}} to örganize" my life.                                                                                                                                                                                                                                                                        \\ 
\hline
\multirow{3}{*}{Conformity}                                 & Obedient                   & 2         & So me and my cousin were in a world and I think there is a bug with parrots with wrong coding because we \textbf{\textit{tamed}} parrots and started hearing monsters sound                                                                                                                                                                                \\ 
\cline{2-4}
                                                            & Self-discipline            & 14        & I love this app but it always crashes and have problems. I used to deal with this, but I have lost my \textbf{\textit{temper}}.                                                                                                                                                                                                           \\ 
\cline{2-4}
                                                            & Politeness                 & 17        &  Incredibly \textbf{\textit{rude}} and unhelpful staff... Calling a customer stupid is not acceptable.                                                                                                                                                                                                                                                                                            \\ 
\hline
\multirow{5}{*}{Tradition}                                  & Respect for tradition      & 7         & If there are already 32 houses on the board and a hotel is sold, 4 new houses will take it is place, which is not how the game \textbf{\textit{traditionally}} works.                                                                                                                                                                                      \\ 
\cline{2-4}
                                                            & Devout                     & 27        & Among other things, it falsely represents that you can use this app to reset your password; you cannot. Plus the support is \textbf{\textit{apathetic}} and generally abysmal.                                                                                                                                                                                                                                   \\ 
\cline{2-4}
                                                            & Detachment                 & 9         & \begin{tabular}[c]{@{}p{13cm}@{}}It was good at first but many are using it for the wrong reason and its putting me off. I'm feeling all these different \textbf{\textit{emotions}} in the span of 5 minutes.\\ \end{tabular}                                                                                         \\ 
\cline{2-4}
                                                            & Humble                     & 3         & \begin{tabular}[c]{@{}l@{}}Poor and tiny food choices served in a very \textbf{\textit{pretentious}} ways in what looked like ashtrays\\ \end{tabular}                                                                                                                                                                                                     \\ 
\cline{2-4}
                                                            & Moderate                   & 8         & In an attempt to look new and \textbf{\textit{edgy}}, the user experience went down the drain. I found it clunky and \textbf{\textit{irritating}}.                                                                                                                                                                                                         \\ 
\hline
\multirow{7}{*}{Benevolence}                                & Helpful                    & 1667      & i wanted to find a hospital parking place in Wythenshawe hospital but this app finds none so Im afraid completely \textbf{\textit{useless}} for me.Im disabled and cannot walk a long way to pay for a parking ticket so i thought the app was going to be very \textbf{\textit{helpful}}.                                                                                   \\ 
\cline{2-4}
                                                            & Responsible                & 29        & Please have a review on "volunteer support" cause some of them do not pay any attention and are \textbf{\textit{irresponsible}}                                                                                                                                                                                                                           \\ 
\cline{2-4}
                                                            & Forgiving                  & 11        & This app is not here to help you it is predatory and waits for you to trust in the trial they have paid forth before \textbf{\textit{punishing}} you for not remembering to end it.                                                                                                                                                                        \\ 
\cline{2-4}
                                                            & Honest                     & 339       & \textbf{\textit{Dishonest}} subscription system that cant be unsubscribed from, charging for parking that never happened on multiple occasions. Taking money for no service is either \textbf{\textit{fraud}} or theft.                                                                                                                                                      \\ 
\cline{2-4}
                                                            & Loyal                      & 4         & \begin{tabular}[c]{@{}l@{}}Get it together... I've been a \textbf{\textit{loyal}} customer for 30 years. But I am considering a swap over for the headaches.\\ \end{tabular}                                                                                                                                                               \\ 
\cline{2-4}
                                                            & A spiritual life           & 4         &  And another thing is that there is no specific way to report \textbf{\textit{religious}} related hate speech.                                                                                                                                                                                                                                             \\ 
\cline{2-4}
                                                            & True friendship            & 4         & Same freezes too often when playing with \textbf{\textit{friends}} online. Have not even completed one game with my friends because of this. Despicable. Good for solo play but why would not you want to destroy \textbf{\textit{friendships}}?                                                                                                                             \\ 
\hline
\multirow{7}{*}{Universalism}                               & Equality                   & 14        & It is really annoying and this is not how the real Monopoly works the randomness does not force you to lose, it supposed to be \textbf{\textit{equal}} chances of winning or losing.                                                                                                                                                                       \\ 
\cline{2-4}
                                                            & Wisdom                     & 50        & \begin{tabular}[c]{@{}p{13cm}@{}}I used this for short and long term goals and notes so I have lost a ton of important stuff. This is what I get for using auto-update on Play store (against my better \textbf{\textit{judgment}}).\\ \end{tabular}                                                                                                             \\ 
\cline{2-4}
                                                            & Inner harmony              & 9         & You are the Number 1 app but you do not \textbf{\textit{cooperate}}!!                                                                                                                                                                                                                                                                                     \\ 
\cline{2-4}
                                                            & A world of beauty          & 39        & New update is \textbf{\textit{ugly}}, the yellow hurts the eyes (even though it is branding)                                                                                                                                                                                                                                                               \\ 
\cline{2-4}
                                                            & Social justice             & 13        & \begin{tabular}[c]{@{}l@{}}Really \textbf{\textit{unfair}} and unrealistic the dice rolls are laughably bad... It seems as though the game is biased...\\ \end{tabular}                                                                                                                                                  \\ 
\cline{2-4}
                                                            & Broadminded                & 7         & There a \textbf{\textit{bias}} against penguin tokens.                                                                                                                                                                                                                                                                                                     \\ 
\cline{2-4}
                                                            & A world at peace           & 93        & There was a minor hiccup a year ago that forced users to use IN or Wipe Key security to continue using the app. While I understand the logic of that decision, it did cause some \textbf{\textit{strife}} for me and other users at the time.                                                                                                             \\
\hline
\end{tabular}
\end{table*}


\begin{table*}
\scriptsize
\centering
\caption{Categories of values violations for 12 apps.}
\label{tab:categoryresult}
\begin{tabular}{p{1.2cm}p{1cm}p{1cm}p{1cm}p{1cm}p{1cm}p{1cm}p{1cm}p{1cm}p{1cm}lp{1.3cm}} 
\hline
              & Achievement & Benevolence & Conformity & Hedonism & Power & Security & Self-direction & Stimulation & Tradition & Universalism & Total \# of Violations  \\ 
\hline
Pinterest     & 21          & 137         & 1          & 184      & 11    & 14       & 200            & 8           & 6         & 32           & 614                     \\
TrainingPeaks & 4           & 58          & 0          & 9        & 1     & 12       & 32             & 2           & 3         & 14           & 135                     \\
Minecraft     & 21          & 89          & 3          & 81       & 3     & 28       & 90             & 5           & 6         & 11           & 337                     \\
Monopoly      & 74          & 158         & 2          & 114      & 69    & 55       & 141            & 13          & 4         & 22           & 652                     \\
PicsArt       & 36          & 236         & 4          & 174      & 6     & 17       & 110            & 4           & 4         & 23           & 614                     \\
Any.do        & 26          & 182         & 1          & 84       & 2     & 16       & 126            & 9           & 1         & 23           & 470                     \\
Telegram      & 19          & 174         & 4          & 68       & 4     & 12       & 121            & 4           & 2         & 16           & 424                     \\
Tripadvisor   & 14          & 152         & 3          & 88       & 4     & 24       & 193            & 8           & 6         & 21           & 513                     \\
PayByPhone    & 37          & 389         & 4          & 88       & 3     & 26       & 191            & 8           & 6         & 27           & 779                     \\
Cellopark     & 22          & 99          & 4          & 25       & 3     & 7        & 54             & 2           & 3         & 18           & 237                     \\
Tiktok        & 16          & 123         & 6          & 115      & 7     & 19       & 89             & 7           & 6         & 23           & 411                     \\
CBA           & 47          & 271         & 1          & 112      & 7     & 35       & 153            & 6           & 7         & 26           & 665                     \\ 
\hline
Average       & 28.1        & 172.3       & 2.8        & 95.2     & 10    & 22.1     & 125            & 6.3         & 4.5       & 21.3         & 487.6                   \\
\hline
\end{tabular}
\end{table*}

\subsection{RQ2: Which reviews indicating human values violations by a mobile app are the most supported by other end users?}

A review is \emph{liked} (i.e., given a thumbs up) if another user considers the review helpful. The more likes there are to a review, the higher the consensus about the issue captured in that review by other users. By extension, it is probable that if such a review contains a values-violation, the other users who support that review might agree with the value consideration captured in the review.

\begin{figure}[!ht]
		\centering 
		\includegraphics[width=0.75\linewidth]{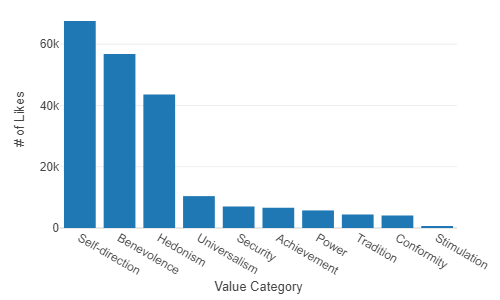}
		\caption{Number of likes for each violated values category.}
		\label{fig:likes}
	\end{figure}

The result of our analysis of the app reviews shows that reviews reporting violations of the human values of Self-direction and Benevolence  garner the most number of likes from other app users. This is consistent with the result in \autoref{sec:reflect} above as shown in \autoref{fig:likes}.

\subsection{RQ3: Are app description features related to specific human values violations?}
Our app feature mining extracts features from user app reviews and supports the identification of features being reviewed. Looking at the features of the apps related to the perceived human values-violations by reviewers can be helpful in isolating the problematic app issues that need to be addressed by app developers.

Our analysis results show how specific app features are associated with different human value violations. For example, the violation of value \emph{pleasure} seems to be strongly related to the app feature \textit{Add stickers} in the Tiktok app. The violation of value item \emph{responsibility} seems to be linked to the app feature \textit{Set reminders} in the Anydo app. \autoref{tab:appfeatures} captures a set of example app features related to human value violations as reported in the analysed app reviews.

\begin{table}
\scriptsize
\centering
\caption{Sample app features and related value-violations.}
\label{tab:appfeatures}
\begin{tabular}{lll} 
\hline
Feature         & App Name      & Value-violation                  \\ 
\hline
Save recipes    & Pinterest     & Curiosity, helpfulness, honesty  \\
Add workouts    & TrainingPeaks & Curiosity, helpfulness           \\
Resource pack   & Minecraft     & Curiosity                        \\
Play online     & Monopoly      & Helpfulness, pleasure            \\
Add stickers    & PicsArt       & Pleasure                         \\
Set reminders   & Any.do        & Responsibility                   \\
Create group    & Telegram      & Helpfulness, Honesty             \\
Add review      & Tripadvisor   & Helpfulness, Honesty             \\
Parking history & PayByPhone    & Curiosity, Helpfulness           \\
Parking payment & Cellopark     & Curiosity                        \\
Watch videos    & Tiktok        & Enjoying life                    \\
Tap \& pay        & CBA           & Helpfulness                      \\
\hline
\end{tabular}
\end{table}

\section{Discussion}
\label{sec:discussion}

\subsection{Implications and Recommendations}
Mobile app development is a significant component of the software industry. Even non-software organisations invest substantial amounts of money into app development and maintenance, whether outsourcing or in-house development, to ensure their organisations can leverage mobile apps. However, as our analysis of app reviews have shown, many popular mobile apps fail to take into account critical human values; this is reflected in the comments left by users in the reviews.

Our work is one of the first steps in detecting perceived violations of human values in mobile software engineering, as our approach was able to detect values violation using over 22,000 app reviews for all 12 apps covering diverse categories and user bases. Given these findings, there is a need to develop new techniques for developing mobile apps to cater to human values considerations and minimise their violations.

Similar to the others in the broader values-oriented software engineering community, we propose extending mobile app requirements engineering and design practices to include the integration of human values consideration as a first class step in software design \cite{Hussain:2018}. For example, by using participatory design methods that are focused on ensuring the values of users are captured and documented as rationales in the requirements phase. Another example is the assignment of a ``critical friend'' in a participatory agile process to ensure design considerations are consistent with the given values of all stakeholders \cite{Whittle:2019}. We advocate raising awareness of the importance of values in mobile software engineering (both in industry and academia) and the merit of values leadership in evolving the software development practice mindset. We also recommend the development of novel methods of validating values in software, because it is one thing to propose values orientation and principles to technical-minded people and quite another to get these principles internalised and reflected in the practices of those for whom they were proposed.

Below we highlight three more possible steps that may be helpful in pursuing our vision of mobile software engineering that not only seeks to address users’ challenges but also solve them in a way that embodies their values.

\subsubsection{Mining Values Requirements}
App reviews are a valuable source of ideas from users and have been mined for various kinds of information, e.g., features requests, bug reports, etc. \cite{Disorbo:2016}. And this information, in turn, has fed into the requirements and incorporated them into future updates of the apps. In a similar approach, key human values requirements for apps can be mined from app reviews. For instance, values violations reflected in the user reviews, although negative, can be seen as values requirements or requests from users. These can be especially helpful when the values requirements are associated with certain app features, e.g., when a user complains about being automatically debited for an app without first receiving a warning signalling the end of the trial period, this can be interpreted as a request for the value of transparency. Future updates of the app can take this into consideration. 

However, we note here that although mining reviews for values requests can be useful, it is still a reactive process, and should be secondary to use of more proactive methods. These might include participatory design methods that ensure that key stakeholders’ values are captured and represented in the first instance.

\subsubsection{Values Alignment}
Ever present is the question of values alignment between stakeholders -- including between software organizations, developers and their end-users -- and how these values are represented in the software artefacts to avoid values violations and conflicts in the first place \cite{Burak:2018}. Since (mobile) software applications are not valueless, this raises the question of whose values are built into the applications \cite{Whittle:2019}. Others have argued that if values are not explicitly captured, the resulting applications will embody the values of their creator \cite{Lennox:2020}, in this case, the values of the software organisations and teams. Thus there is a key need for a proactive values alignment discussion beyond values-violation detection.

One of the resulting principles of the 2017 Asilomar AI principles states that ``highly autonomous AI systems should be designed so that their goals and behaviours can be assured to align with human values throughout their operation'' \cite{Asilomar:2020}. We affirm and extend this principle to mobile software engineering practice, as mobile applications are both ubiquitous and are vehicles of AI models.

There is also the challenge of matching the values of stakeholders, and resolving values conflict when there is a disagreement between values representations, e.g., between the values embodied in the software artefact and the users’ expectations, or between the values espoused by developers and those of the users.

While reviews analysis such as the one discussed in this work can be carried out to understand areas of values-violations in mobile apps, nothing can substitute the place of a proactive values alignment discussion and agreement with all stakeholders.

\subsubsection{Towards a Critical Technical Practice in Mobile Software Engineering}
Smartphones, first released in 2007, have affected society in many ways. Mobile applications built for smartphones are used for many daily tasks, both ordinary and complex; from deciding our route of daily commutes, to many work practices, to supporting our physical activities. It is essential to understand the sociocultural context of technologies such as mobile applications and how they affect or change society at large; acknowledging the harmony that ought to exist between mobile technologies and users’ values system.

Understanding the space between the technical aspects of mobile software development and the reflective work of sociocultural criticisms that highlight the hidden assumptions in the technical process is what a critical technical practice is about, and has been embraced in AI \cite{Agre:1997}, HCI \cite{Dourish:2004}, and design \cite{Sengers:2006}. An ongoing critical technical practice would be invaluable in making these assumptions and embedded value systems precise and visible, and support their enquiry in relation to the field of mobile software engineering.

As shown in recent research and events, e.g., \cite{ACS1:2020,Mohamed:2020}, strictly technical approaches fail to address ethical and external factors of systems \cite{Mohamed:2020}, thus failing to reflect the value considerations of stakeholders. A critical technical practice would encourage the positive interaction between the social, cultural and technical aspects of mobile software artefacts. This would involve taking a step outside the technical field of specialisation, i.e., mobile software engineering, and would limit the challenges of trying to apply technical mindsets to non-technical problems \cite{Agre:1997}. 

A critical technical practice will entail engaging with the social and human sciences approaches such as critical theory \cite{Agre:1997} to open up the assumptions underpinning mobile software engineering and how they affect the social and cultural aspects of society, and in so doing support the evaluation of the mobile software engineering field’s contribution to society.

\subsection{Current Limitations}
A current limitation of our approach is its inadequacy to detect values violations in languages other than English. Because our approach utilises a manually crafted values-dictionary of English words, we cannot apply our approach to app reviews written in other languages without repeating the values-dictionary creation process for a specific language. Although it might be helpful to use automatic translators to translate other languages to English before applying our technique, there are many aspects, nuances and local accents that may very well be lost in translation \cite{Shams:2020}. Our future work in this regard will be to continue our work on labelling different datasets and adopt a machine learning or hybrid approach that is generalisable.

A second limitation of our work is our exclusion of 8 value items from our values-dictionary resulting in our technique being able to detect only 50 (out of a total of 58) value items from the Schwartz’ model of basic human values. The excluded value items include the following: social power, reciprocation of favours, honouring of parents and elders, accepting my portion in life, mature love, meaning in life, unity with nature, and protecting the environment. These were excluded because they consist of phrases whose keywords and associated synonyms and antonyms could not be approximated in our values-dictionary. While a value item with the phrase, “preserving my public image”, could be approximated to a similar word such as “reputation”, we could not approximate similar words and their associated synonyms and antonyms for the excluded value items. As already mentioned above, it would be worthwhile to apply a machine learning or hybrid approach, as more labelled datasets become available.

\section{Related Work}
\label{sec:relatedwork}

\vspace{0.1cm}\noindent{\bf Mining App User Feedback. }
Several researchers have carried out work on mining app reviews to understand feedback provided by users in the form of reviews and to help developers with this useful information \cite{Harman:2012,Chen:2014,Khalid:2015}. Using natural language processing (NLP) techniques, Guzman and Maalej analysed app reviews and extracted fine-grained app features that developers found useful in requirements evolution tasks \cite{Guzman:2014}. A similar work employed predefined linguistic rules and Latent Dirichlet Allocation (LDA) techniques to identify and retrieve feature requests from app reviews, showing amongst other results, that users typically request for better support and more frequent updates to mobile apps \cite{Iacob:2013}. 

To support software requirements evolution, Li et al. introduced an evaluation framework for understanding user satisfaction from user comments \cite{Li:2010}. Another study applied sentiment analysis and topic modelling to extract important topics useful for requirements from reviews \cite{Carreno:2013}. To ease the effort required in analysing app reviews, Di Sorbo et al. proposed SURF, a tool for summarising large number of reviews into coherent summaries and recommends informative software changes \cite{Disorbo:2016}.
Other studies have focused on automatically classifying app reviews. For example, Panichella et al. proposed a taxonomy for classifying reviews and introduced a hybrid approach of NLP, text analysis, and sentiment analysis to classify app reviews into their proposed taxonomy \cite{Panichella:2015}, while Maalej and Nabil utilised probabilistic techniques to classify reviews into four categories: bug reports, feature requests, user experiences, and ratings \cite{Maalej:2015}.

Our work is complementary to the studies highlighted above, as we also aim to understand feedback provided by users in app reviews, albeit from a human values perspective. Our work focuses on a different aspect - the detection of human values violation in app reviews.

\vspace{0.1cm}\noindent{\bf Values-based Software Engineering.}
Values-based software engineering (SE) is a relatively nascent and growing area of SE and has begun to receive attention from the SE research community in recent years \cite{Mougouei:2018}. A recent study of the prevalence of human values in SE publications show that only a small portion of SE publications directly consider human values, and within these publications, the majority of values are largely neglected \cite{Perera:2020}.

Research into the study of human values in the social sciences (e.g., Schwartz’s theory of basic human values \cite{Schwartz:2012}) has contributed to values-based design and values-based SE \cite{Perera:2020}. Studies by Friedman highlight the relationship between technological tools and ethics, morals, and values, and call for values-sensitive design - a principled manner through which technology can account for values in the design process \cite{Friedman:1996,Friedman:2017}. While Friedman’s value-sensitive design “emphasises values with an ethical import”, others have argued for the study of the role of all human values in SE, beyond an ethically oriented subset, with the aim of understanding the interdependence relationship between different values \cite{Winter:2018}. Using real-world projects with non-profits, Whittle et al. made a case for human values in SE based on Schwartz’s value model, and how they can be integrated into existing SE practices \cite{Whittle:2019}.

Moreover, Ferrario et al. proposed the concept of values-first SE based on Schwartz’s value model, specifically in the decision-making process, and applied these principles to a case study in the health domain \cite{Ferrario:2016}. Building on the values-first SE concept, Winter et al.  introduced the Values Q-sort -  a values measurement tool for investigating values at different levels of abstraction in SE \cite{Winter:2018}.
Some studies have proposed techniques for values-based SE, e.g, a conceptual model for values-sensitive design \cite{Barn:2015}, values-design hub \cite{Hussain:2018}, while others studies have mapped human values to other ethical principles, e.g., mapping principles and rights of GDPR to human values \cite{Perera:2019}, mapping human values to the ACM code of ethics \cite{Winter:2018}.

More recently, Shams et al. conducted a case study to understand the desired and missing human values in existing Bangladeshi agriculture mobile apps by manually analysing reviews from these apps based on the Schwartz’s values model \cite{Shams:2020}. To the best of our knowledge, only the study by Shams et al. is closely related to our work, as it is focused on mobile apps reviews. However, our work utilises automatic values-violation detection -  a generic NLP approach to detecting the violation of human values and associated app features in app reviews, and can be applied to a broad category of mobile apps.

\section{Conclusions}
\label{sec:conclusion}

Consideration of human values in the development of software systems is essential, as ubiquitous software such as mobile apps can have a very significant positive or negative impact on individuals and society at large. Human values can serve as lenses through which we evaluate and mitigate the dangers associated with technology on modern society.
In this paper, we analysed app reviews to detect perceived end user human values violations in 12 common mobile apps using NLP techniques. Our approach utilises a combination of sentiment analysis, app features extraction, and a dictionary of human values keywords based on the widely accepted Schwartz theory of basic human values. Our results show that many perceived violations of human values are reflected in apps reviews -- a quarter of the reviews analysed contained indicators of human values violations. We found that benevolence and self-direction were the most violated value categories, and that reviews reporting these also garnered the most support (in the form of likes) from other users.

The implications of our results recognise the need for a proactive approach to the elicitation and alignment of human values amongst stakeholders involved in the development of mobile apps to prevent conflicts and values violation. Moreover, app reviews can also serve as a valuable source for mining end user human values requirements to support the evolution of mobile apps via updates and features. We advocate a move towards a critical technical practice in mobile software engineering of employing critical methods and socio-cultural reflection of the roles and effects of mobile apps in our society.

\bibliographystyle{IEEEtran}




\end{document}